\pdfoutput=1

\documentclass[11pt]{article}
\usepackage{moriond,epsfig}

\usepackage{amssymb}
\usepackage{amsmath,amsbsy}
\usepackage{graphicx}
\usepackage{epsfig}
\usepackage{color}
\usepackage{cite}

\def\MyPhoto{\includegraphics[height=60mm]{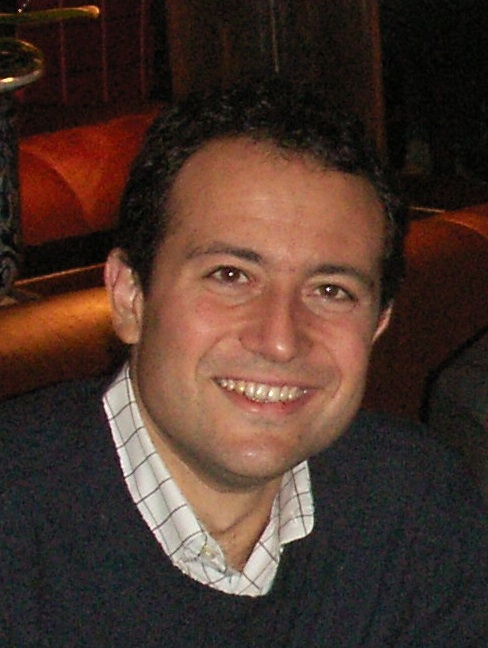} \includegraphics[height=60mm]{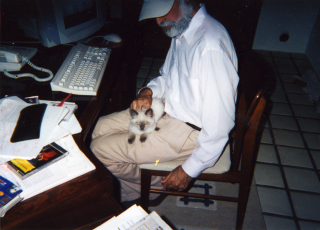}}

\def\beq{\begin{equation}}
\def\eeq{\end{equation}}
\def\bea{\begin{eqnarray}}
\def\eea{\end{eqnarray}}

\def\l{\lambda}
\def\gtwid{{\,\raise.35ex\hbox{$>$\kern-.75em\lower1ex\hbox{$\sim$}}\,}}
\def\ltwid{{\,\raise.35ex\hbox{$<$\kern-.75em\lower1ex\hbox{$\sim$}}\,}}

\def\mev{{\rm MeV}}

\begin{document}
\vspace*{4cm}
\title{Demise of CKM \& its aftermath~\footnote{Invited
talk at the EW Moriond 2010}}.

\author{Enrico Lunghi$^\dagger$ and Amarjiit Soni$^{\dagger\dagger}$\footnote{Speaker}}
\address{${}^{\dagger}$ Physics Department, Indiana University, Bloomington, IN 47405, USA \\
${}^{\dagger \dagger}$ Department of Physics, Brookhaven National Laboratory,Upton, NY 11973, USA}

\maketitle\abstracts{Using  firmly established experimental inputs such as $\epsilon_K$, $\Delta M_d$, $\Delta M_s$, Br($B \to \tau \nu$), $\gamma$, $V_{cb}$ along with corresponding lattice matrix elements which have been well studied and are in full QCD such as $B_K$, SU3 breaking ratio $\xi$, $B_{B_s}$ and in particular without using $V_{ub}$ or the pseudoscalar decay constants $f_{B_d}$ or $f_{B_s}$ from the lattice, we show  
that the CKM-paradigm now appears to be in serious conflict with the data. Specifically the SM predicted value of $\sin 2 \beta$ seems too high compared to direct experimental measured value by over 3$\sigma$. Furthermore, our study shows that new physics predominantly effects B-mixings and $B_d \to \psi K_s$, and not primarily in kaon-mixing or in $B \to \tau \nu$. Model independent operator analysis suggests the scale of underlying new physics, accompanied by a BSM CP-odd phase, responsible for breaking of the SM is less than a few TeV, possibly as low as a few hundred GeV. Two possible BSM scenarios, namely warped extra-dimensions and SM with a 4th generation, are briefly discussed. Generic predictions of warped flavor models are briefly discussed. While SM with 4th generation (SM4) is a very simple way
to account for the observed anomalies, SM4 is also well motivated
due to its potential role in dynamical electroweak symmetry breaking 
via condensation of heavy quarks and in baryogenesis.
}

\section{\bf Introduction}

The next big step in our understanding of particle physics will be the uncovering of the electro-weak symmetry breaking (EWSB) mechanism. The present and upcoming collider experiments (Fermilab and LHC) will be able to test the Standard Model (SM) Higgs mechanism. New physics is widely expected at around the TeV scale if the Higgs mass is not to receive large radiative corrections and require severe  fine-tuning. A stringent constraint on the SM mechanism of EWSB is the tight structure of flavor changing (FC) interactions: tree--level FC neutral currents are forbidden and charged currents are controlled by the Cabibbo--Kobayashi--Maskawa (CKM)~\cite{ckm} mixing matrix
\beq
V = 
\begin{pmatrix}
1-\frac{\l^2}{2} & \l & A \l^3 (\rho- i \eta)\cr
-\l & 1-\frac{\l^2}{2} & A \l^2 \cr
A \l^3 (1-\rho-i \eta) & - A \l^2 & 1 \cr
\end{pmatrix} .
\eeq
Within the SM, the CKM matrix is the only source of FC interactions and of CP violation. There is no reason, in general, to expect that new physics (needed to stabilize the Higgs mass) at the TeV scale will be in the basis wherein the quark mass matrix is diagonal. This reasoning gives rise to another fundamental problem in particle physics, namely the flavor puzzle i.e. unless the scale of new physics is larger than $10^3$ TeV it causes large FCNC especially for the $K-\bar K$ system. Thus flavor physics provides constraints on models of new physics up to scales that are much much larger than what is accessible to direct searches at colliders such as  the Tevatron or the LHC. Flavor physics is therefore expected to continue to provide crucial information for the interpretation of any physics that LHC may find.

In the past  decade significant progress was made in our understanding of flavor physics, thanks in large part to the spectacular performance of the two asymmetric B-factories. For the first time it was experimentally established that the CKM-paradigm~\cite{ckm} of the Standard Model (SM) provides a quantitative description of the observed CP violation, simultaneously in the B-system as well as in the K-system with a single CP-odd phase, to an accuracy of about 20\%~\cite{Nir02}.  While this success of the CKM picture is very impressive, the flip side is that an accuracy of $O(20\%)$ leaves open the possibility of quite sizable new physics contributions. In this context it is important to recall that the indirect CP violation parameter, $\varepsilon_K \sim 2 \times 10^{-3}$~\cite{PDG_10} is an asymmetry of $O(10^{-3})$ and an important reminder that if searches had been abandoned even at $O(1\%)$ the history of Particle Physics would have been completely different. Indeed, in the past few years as better data and better theoretical calculations became available some rather serious tensions have emerged~\cite{Lunghi:2007ak,Lunghi:2008aa,Bona:2009cj,Lunghi:2009sm,Lunghi:2009ke,Lenz:2010gu}.

Recently~\cite{Lunghi:2010gv},  we showed that the use of the latest experimental inputs along with a careful use of the latest lattice results leads to a rather strong case  for a sizable contribution due to beyond the Standard Model sources of CP violation that in $\sin 2 \beta$ could be around 15-25\%. Clearly if this result stands further scrutiny it would have widespread and significant repercussions for experiments at the intensity as well as the high energy frontier.  We also were able to isolate  the presence of new physics primarily in the time dependent CP measurements via the ``gold-plated" $\psi K_s$ mode which intimately involves B-mixing amplitude and the decay $B \to \psi K_s$. Our analysis does not exclude possible sub-dominant effect in kaon-mixing and/or in $B \to \tau \nu$. In particular, our analysis~\cite{Lunghi:2010gv} indicates  that the data does not seem to provide a consistent interpretation for the presence of large new physics contribution to the tree amplitude for $B\to \tau \nu$. 

\section{\bf Choice of lattice inputs}
Key inputs from experiment and from the lattice needed for our analysis are shown in Table~\ref{tab:utinputs}. Below we briefly remark on a few of the items here that deserve special mention: \\

\noindent $\bullet$ With regard to lattice inputs we want to emphasize that quantities used here have been carefully chosen and  are extensively studied on the lattice for many years. Results that we use are obtained in full QCD with $N_F= 2 + 1$ simulations, are in the continuum limit, are fairly robust and emerge from the works of more than one collaboration and in most cases by many collaborations. \\
\noindent $\bullet$ Regarding calculations of $\hat B_K$~\cite{BK_define} on the lattice, it is useful to note that in the past 3 years a dramatic reduction in errors  has been achieved and by now a number of independent calculations with errors $\ltwid 5\%$ and with consistent central values have been obtained  using $N_f = 2 + 1$~\cite{Gamiz:2006sq,RBC-UKQCD07,Aubin:2009jh,Kim:2009te,Bae:2010ki} as well as $N_f = 2$~\cite{Aoki:2008ss} dynamical simulations (see Ref.~\cite{Laiho:2010conf} for a review). Again, to be conservative, we only use weighted average of two results that are both in full QCD, use different fermion discretizations (domain-wall and staggered) and that also use completely different gauge configurations and completely different procedures for operator renormalization~\cite{Aoki:2010pe,Bae:2010ki}. 
\begin{figure}[t]
\begin{center}
\includegraphics[width=0.7 \linewidth]{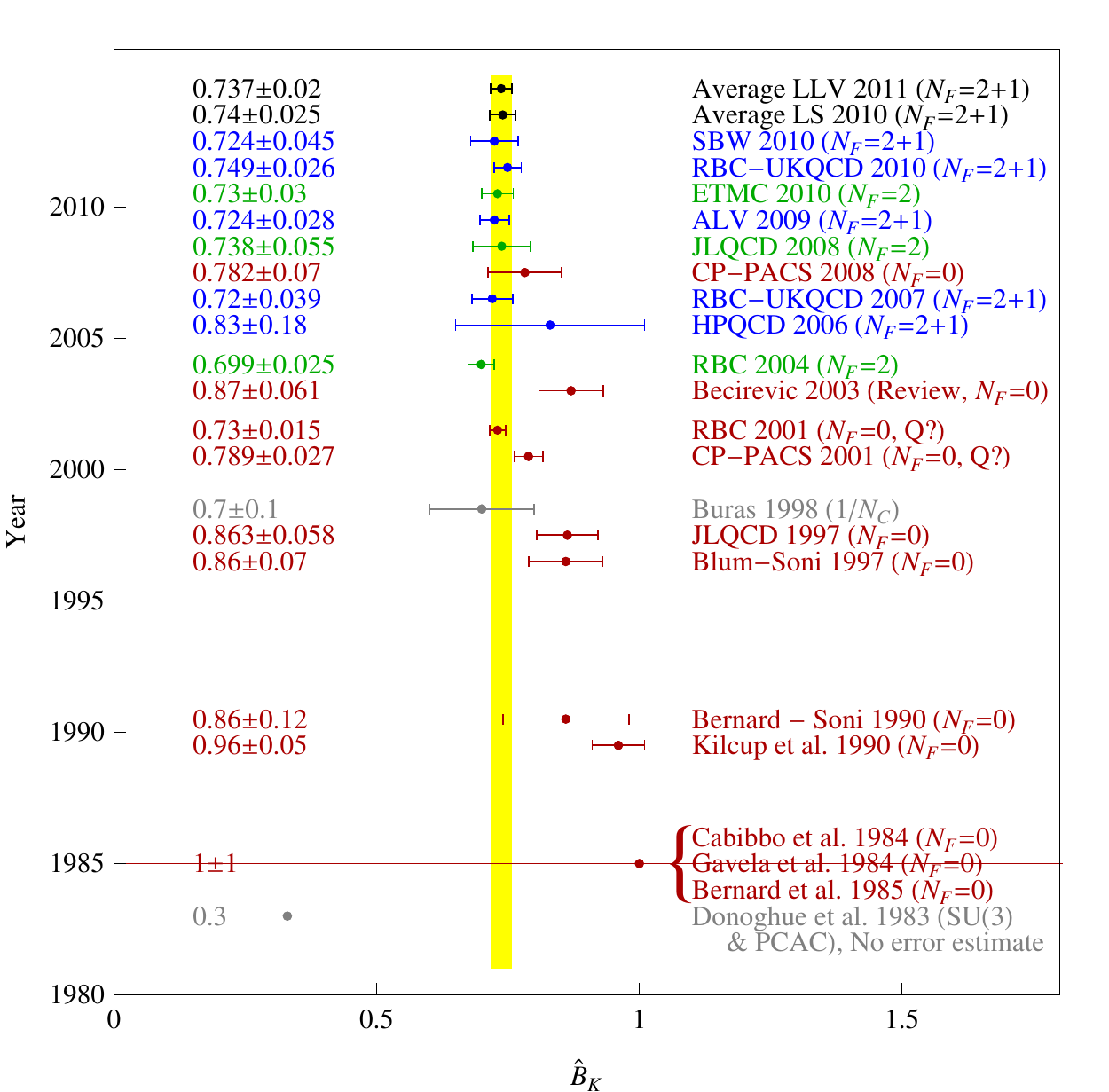}
\caption{A quarter century of lattice--QCD efforts to improve the determination  of $B_K$. As a representative of continuum methods, the results of Refs.~\cite{Donoghue:1982cq,Ginsparg:1983kf} using lowest order ChPT and $SU(3)$ flavor symmetry with no estimate of errors and of Ref.~\cite{Bardeen:1987vg,Buras:1998raa} using large $N_C$ are shown in gray. Several quenched lattice results are shown in red. In particular, three earliest attempts on the lattice around 1984-85 (marked as Cabibbo {\it et al.}, Gavela {\it et al.} and Bernard {\it et al.})~\cite{Cabibbo:1983xa, Brower:1984ta, Bernard:1985tm}. Amongst these early attempts are also ~\cite{Bernard:1989mv} using Wilson Fermions and of~\cite{Kilcup:1989fq} using staggered quarks. First large scale staggered result, marked as JLQCD is of~\cite{Aoki:1997nr}. Blum and Soni~\cite{Blum:1997mz} marks the first simulation of $B_K$ using quenched domain-wall quarks followed by large scale studies of that approach by CP-PACS 2001~\cite{AliKhan:2001wr}, RBC 2001~\cite{Blum:2001xb} and CP-PACS 2008~\cite{Nakamura:2008xz}. Review of all quenched results prior to 2003 is marked as Becirevic~\cite{becirevic}. Unquenched 2 flavors calculations are shown in green including RBC 2004~\cite{Dawson:2004gu}; JLQCD 2008~\cite{Aoki:2008ss} and  ETMC 2009~\cite{Bertone:2009bu,Constantinou:2010qv}. Unquenched 2+1 flavors calculations are shown in Blue: HPQCD 2006~\cite{Gamiz:2006sq} with staggered; RBC-UKQCD 2007~\cite{Cohen:2007zz} with domain wall and  ALV~\cite{Aubin:2009jh} using mixed action. Also RBC-UKQCD 2010~\cite{Aoki:2010pe} using domain-wall quarks and  SBW~\cite{Bae:2010ki} using staggered. In black [LS]  is the average used in the analysis of Ref.~\cite{Lunghi:2010gv} and the recent average[LLV] obtained in Refs.~\cite{Laiho:2009eu,Laiho:2011nz}. \hfill
\label{fig:BKhistory}}
\end{center}
\end{figure}

Given the important role lattice calculations of weak matrix element are playing in the evaluation of some of the important non-perturbative quantities that are critical to constraining fundamental parameters in flavor physics,  we now take this opportunity to briefly comment on how the calculation of $B_K$ evolved over the past $\sim \!\! 25$ years. This example should serve to illustrate developments in many such calculations; the history of $B_K$ is summarized in Fig.~\ref{fig:BKhistory}~\cite{thanks_Taku}.

Recall that before the advent of the lattice approach to $B_K$, an interesting first attempt~\cite{Donoghue:1982cq,Ginsparg:1983kf}, using charged kaon lifetime, lowest order chiral perturbation theory and flavor SU(3) symmetry, estimated $B_K \approx 0.33$. But of course an estimate such as this represents an uncontrollable approximation, with no reliable error estimate or understanding of scale dependence. If one were to use such a value of $B_K$, in conjunction with experimentally measured value of $\epsilon_K$, to deduce the Wolfenstein parameter $\eta$, that uniquely controls CP violation in the CKM picture, we would get a central value about a factor of two higher than modern numbers, but even more noticeably the error on $\eta$ could easily be O(100\%) rather than $\sim 10\%$~\cite{Lunghi:2009sm} that we now have.                                                          

One of the primary purpose for the construction of the two asymmetric B-factories in the 1990's was that they would allow us to extract directly from experiment the weak CP-odd phase via $B \to \psi K_S$. They accomplished this task beautifully, providing us with a rather precise value, $\sin 2 \beta = 0.668 \pm 0.023$, {\it i.e.} with an accuracy of about 3.4\%. But for this accurate determination, attained at a significant expense and effort, to be useful in testing the Standard Model, and in particular to test that the CP-odd phase in the CKM paradigm  is quantitatively responsible for the observed CP violation in $K_L$ decays as well as in B-decays, a value for $B_K$ with commensurate precision is essential. If the accuracy on $B_K$ had stayed at the level of O(100\%) then the B-factory measurement would have failed to have an impact on the fundamental theory. 

The very first attempts~\cite{Cabibbo:1983xa,Bernard:1985tm, Brower:1984ta} in the 80's on calculating $B_K$ all started with Wilson fermions in the quenched approximation and of course had huge errors with a value of $B_K$ consistent with 0 or 1.

Amongst the continuum methods,  perhaps the most interesting result was that of~\cite{Bardeen:1987vg}, $\hat{B}_K = 0.70 \pm 0.10$~\cite{Buras:1998raa}, obtained by using the large N approximation: this result corresponds to lattice results obtained in the quenched approximation. Remarkably, this calculation has been consistent with all older lattice results obtained over the years in the quenched approximation, and in fact its claimed accuracy is higher, since, following~\cite{Sharpe:1996ih} many, if not most, lattice calculations done  in the quenched approximation were quoting a (conservative) guess-estimate for the systematic error due to the quenched approximation of $\approx 15\%$~\cite{Becirevic:2004fw}, though with hindsight we now see that the quenching error on $B_K$ was less than 5\%.

In the quenched approximation, the result, $\hat {B}_K = 0.863 \pm 0.058$ (where the stated error does not include quenching error) of~\cite{Aoki:1997nr}, obtained by using staggered quarks, and perturbative renormalization, stood as the most precise lattice result for a long time. With the advent of domain-wall quarks~\cite{ShamirFurman,BlumS1,Blum:1997mz} and with large scale (quenched) simulations with domain wall quarks~\cite{AliKhan:2001wr,Blum:2001xb} it was found that domain wall quarks consistently tended to give about 8 to 15\% smaller $B_K$ (implying a larger CP-odd phase, $\eta$) compared to the staggered result of~\cite{Aoki:1997nr}.   

With dynamical 2+1 simulations there is no longer any need for estimating quenching errors and 3-4 years ago RBC--UKQCD~\cite{RBC-UKQCD07} obtained the first result in full QCD using DWQ, with an estimated total  error of about 5.5\%, finally managing to by-pass the stated accuracy of ~\cite{Buras:1998raa}. Furthermore, by 2010 quite a few lattice calculations using full QCD (and with different discretizations~\cite{Laiho:2010conf}) have managed to reduce the error even less than that to about 4\%. Furthermore, the latest RBC-UKQCD calculation~\cite{Aoki:2010pe} made significant improvements in renormalization and in chiral extrapolation to reduce the  error further to 3.6\%.

As mentioned previously, in our study we are using a weighted average of this latest domain-wall result~\cite{Aoki:2010pe} and that of Ref.~\cite{Bae:2010ki} obtained by using staggered quarks. \\

\noindent $\bullet$ Given the large disparity between the exclusive and inclusive determinations of $V_{ub}$ at the level of $3.3\sigma$, see Table \ref{tab:utinputs}, it is very difficult to draw  reliable conclusions by using this quantity; therefore, since 2008~\cite{Lunghi:2008aa} we have been advocating not using $V_{ub}$ for testing or constraining the UT. Consequently in this work also we will make very limited and peripheral use of $V_{ub}$ only. We should also stress that this is one of the key differences between other groups~\cite{Lenz:2010gu,Bona:2009cj} work on UT fits and us. Another difference from those works is that we use $\xi = f_{B_s} \hat B_s^{1/2}/f_{B_d} \hat B_d^{1/2}$, $f_{B_s} \hat B_s^{1/2}$ and $\hat B_d$ to describe $B_q$ mixing and $B\to \tau\nu$ (as opposed to $f_{B_s}/f_{B_d}$, $\hat B_s/\hat B_d$, $\hat B_s$ and $f_{B_s}$). Moreover, we fit $f_{B_d}$ in conjunction with particular hypotheses and use the determined value of $f_{B_d}$ as a diagnostic tool. Another difference between our work and Ref.~\cite{Lenz:2010gu} is that, in the latter, the authors include in the fit the D0 dimuon asymmetry~\cite{Abazov:2010hv,Abazov:2010hj}, $a_{SL}^s$, in semileptonic $B_s$ and $\bar B_s$ decays; while we agree that this is a very interesting result, we believe that it needs confirmation and therefore are not using it in our fit.\\

\noindent $\bullet$ The complete set of lattice inputs that we use is presented in Table~\ref{tab:utinputs}. All inputs, are taken from 
Refs.~\cite{Laiho:2009eu, Laiho:2011nz} (see \texttt{http://www.latticeaverages.org} for updates) with the exception of $\hat B_K$ (see discussion above), $\xi$ (since the statistical errors of the HPQCD and Fermilab/MILC results are 100\% correlated, we decided to increase the statistical error of the HPQCD result to bring it in line with the with the more conservative Fermilab/MILC estimate), $f_{B_d}$ (we update the HPQCD determination of $f_{B_d}$~\cite{fb}) and $f_{B_s} \hat B_s^{1/2}$ (we update the HPQCD determination of $f_{B_s}$~\cite{fb} and combine it with the Fermilab/MILC result; we then combine the $f_{B_s}$ average with the HPQCD determination of $\hat B_s$ adding {\it linearly} the uncertainties.). 
\begin{table}[t]
\begin{center} 
\begin{tabular}{|l|l|} \hline
$\left| V_{cb} \right|_{\rm excl} =(39.5 \pm 1.0) \times 10^{-3}$& 
$\eta_1 = 1.51 \pm 0.24$~\cite{Herrlich:1993yv} \cr
$\left| V_{cb} \right|_{\rm incl} =(41.68 \pm 0.44 \pm 0.09 \pm 0.58) \times 10^{-3}$~\cite{HFAG10}&  
$\eta_2 = 0.5765 \pm 0.0065$~\cite{Buras:1990fn} \cr
$\left| V_{cb} \right|_{\rm avg} =(40.9 \pm 1.0) \times 10^{-3}$  & 
$\eta_3 = 0.494 \pm 0.046$\cite{Herrlich:1995hh,Brod:2010mj} \cr
$\left| V_{ub} \right|_{\rm excl} = (31.2 \pm 2.6) \times 10^{-4} $  &
$\eta_B = 0.551 \pm 0.007$~\cite{Buchalla:1996vs}  \cr
$\left| V_{ub} \right|_{\rm incl} = (43.4 \pm 1.6^{+1.5}_ {-2.2}) \times 10^{-4} $~\cite{HFAG10}     & 
$\xi  = 1.23 \pm 0.04$ \cr
$\left| V_{ub} \right|_{\rm tot} = (33.7 \pm 4.9) 10^{-4}$  & 
$\lambda = 0.2253  \pm 0.0009$~\cite{Antonelli:2010yf}\cr

$\Delta m_{B_d} = (0.507 \pm 0.005)\; {\rm ps}^{-1}$ & 
$\alpha = (89.5 \pm 4.3)^{\rm o}$\cr
$\Delta m_{B_s} = (17.77 \pm 0.12 )\;  {\rm ps}^{-1}$ & 
$\kappa_\varepsilon = 0.94 \pm 0.02$~\cite{Buras:2008nn,Laiho:2009eu,Buras:2010pza} \cr
 $\varepsilon_K = (2.229 \pm 0.012 ) 10^{-3}$ &
 $\hat B_d = 1.26 \pm 0.11$\cr
$m_{t, pole} = (172.4 \pm 1.2) \; {\rm GeV}$ &
$f_{B_d} = (208 \pm 8) \; {\rm MeV}$~\cite{fb} \cr
$m_c(m_c) = (1.268 \pm 0.009 ) \; {\rm GeV}$ &
$f_K = (155.8 \pm 1.7) \; {\rm MeV}$ \cr
$S_{\psi K_S} = 0.668 \pm 0.023$~\cite{MK_10} &
$\hat B_K = 0.742 \pm 0.023$\cr
$f_{B_s} \sqrt{\hat B_{B_s}} = (291 \pm 16) \; \mev $ &
$\gamma = (78 \pm 12)^{\rm o}$~\cite{Bona:2005vz,Bona:2006ah} \cr
${\rm BR}_{B\to \tau\nu} = (1.68\pm 0.31) \times 10^{-4}$~\cite{Ikado:2006un,BaBar:2010rt,Hara:2010dk} &
\cr
\hline
\end{tabular}
\caption{Lattice QCD and other inputs to the unitarity triangle analysis. The determination of $\alpha$ is obtained from a combined isospin analysis of $B\to (\pi\pi,\; \rho\rho, \; \rho\pi)$ branching ratios and CP asymmetries~\cite{HFAG10}. Statistical and systematic errors are combined in quadrature; for the error on $V_{ub}$ see~\cite{footnote_Vub}. We adopt the averages of Ref.~\cite{Laiho:2009eu, Laiho:2011nz} (updates at  \texttt{www.latticeaverages.org}) for all quantities with the exception of $\xi$, $f_{B_s} \hat B_s^{1/2}$, $\hat B_K$ and $f_{B_d}$ (see text).\hfill
\label{tab:utinputs}}
\end{center}
\end{table}
\section{\bf Some results of the fit.}
\begin{figure}[t]
\begin{center}
\includegraphics[width=0.6 \linewidth]{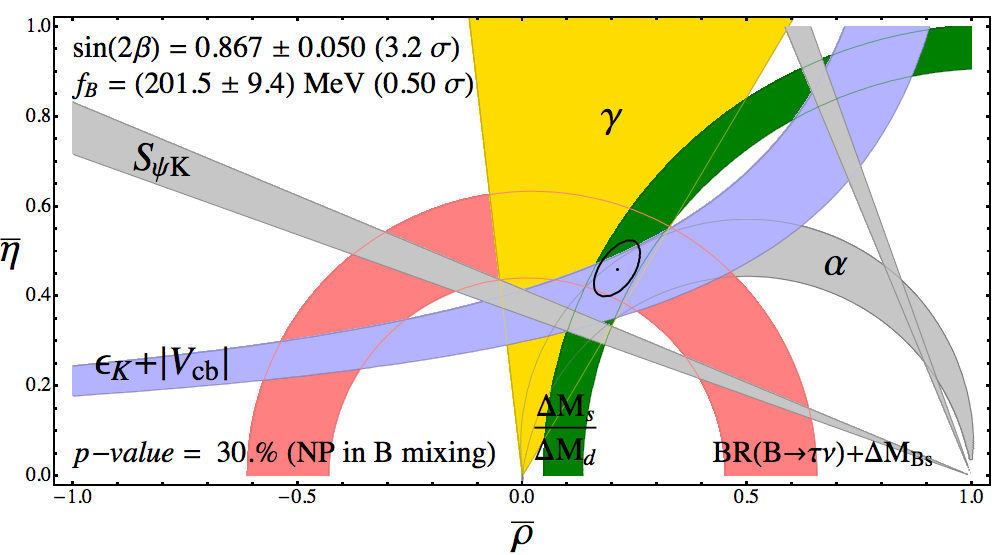}
\includegraphics[width=0.6 \linewidth]{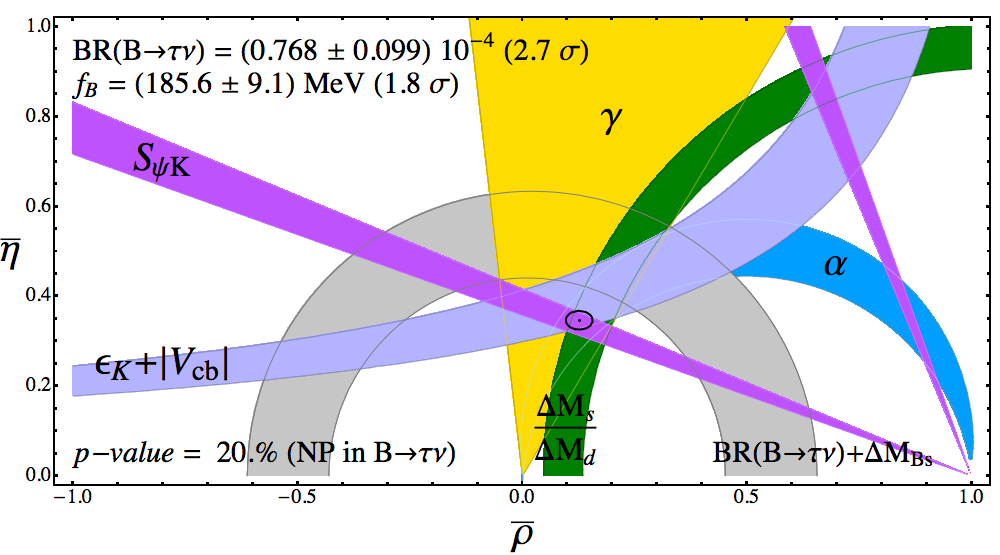}
\caption{Unitarity triangle fit. In each plot inputs that are grayed out are {\it not} used to obtain the black contour (which represents the SM allowed $1\sigma$ region), the p--value and the fit predictions presented in the upper left corners. The deviations of the fit predictions for $\sin(2\beta)$ and ${\rm BR}(B\to\tau\nu)$ from the respective measurements are obtained using the actual chi-square distribution for these quantities. The p-value of the complete SM fit (i.e. including all the inputs) is $p_{\rm SM} = 1.7\%$. In the upper panel, we consider a scenario with a new phase in $B_d$ mixing, thereby removing the $\sin(2\beta)$ and $\alpha$ inputs. In the lower panel we consider a scenario with new physics in $B\to \tau \nu$, thereby removing the ${\rm BR}(B\to\tau\nu)$ input. \hfill
\label{fig:utfit}}
\end{center}
\end{figure}

We first draw attention to the results of the fit shown in the upper panel of Fig.~\ref{fig:utfit}. Here we use as inputs from experiments, $\epsilon_K$, $\Delta M_d$, $\Delta M_s$, $\gamma$ and ${\rm BR} (B \to \tau \nu)$~\cite{no_alpha} and from the lattice, $\hat B_K$, $\xi$, $f_{B_s} \hat B_s^1/2$ and $\hat B_d$ (but not $f_{B_d}$) and we extract the fitted value of $\sin 2 \beta$ and of $f_{B_d}$. We obtain:
\begin{equation}
\sin(2\beta)^{\rm fit} = 0.867 \pm 0.050 \; , \label{sin2betafit}
\end{equation}
which is about 3.2 $\sigma$ away from the experimentally measured value of $0.668 \pm 0.023$. We believe this result  provides  a strong indication that the CKM description of the observed CP violation is breaking down~\cite{no_gamma}.

For the fitted value of $f_{B_d}$ along with the predicted value of $\sin (2 \beta)$ given above, we find:
\beq
f_{B_d}^{\rm fit} = (201.5 \pm 9.4)\; {\rm MeV} \; . \label{fbsin2beta}
\eeq
This ``predicted" value is in very good agreement with the one obtained by direct lattice calculation, $f_{B_d} = (208 \pm 8)$ MeV. This is a useful consistency check signifying that the SM description of the inputs used, especially of $B \to \tau \nu$, is working fairly well and that it is unlikely that the $B\to \tau\nu$ tree amplitude is receiving large contributions from new physics; most likely the dominant effect of new physics is in fact in $\sin (2 \beta)$. Later we will reexamine this from an entirely different perspective and show in fact there is additional independent support to these interpretations. 

In order to  further scrutinize the tentative conclusion reached above, we next present an alternate scenario depicted in the bottom panel of Fig.~\ref{fig:utfit}. Here, we make one important change in the inputs used. Instead of using the measured value of ${\rm BR} (B \to \tau \nu)$ we now use as input the measured value of $\sin (2 \beta)$ from the ``gold-plated" $B_d \to \psi K_s$ mode. Again, this fit yields two important predictions:
\begin{align}
{\rm BR} (B\to \tau\nu)^{\rm fit} &=(0.768\pm 0.099) \times 10^{-4} \; , \label{btnfit} \\
f_{B_d}^{\rm fit} &= (185.6 \pm 9.1)\; {\rm MeV} \; . \label{fbbtn}
\end{align}
Eq.~(\ref{btnfit}) deviates by $2.7\sigma$ from the experimental measurement, as can also be gleaned from an inspection of the bottom panel of Fig.~\ref{fig:utfit}. It is particularly interesting that also the fit prediction for $f_{B_d}$ now deviates by about $1.8\sigma$ from the direct lattice determination given in Table~\ref{tab:utinputs}. We believe this provides additional support that the measured value of $\sin (2 \beta)$ being used here as a key input is not consistent with the SM and in fact is receiving appreciable contributions from new physics.

This leads us to conclude that while the presence of some sub-dominant contribution of new physics in $B\to\tau\nu$ is possible, a large contribution of new physics in there is not able to explain, in a consistent fashion, the tension we are observing in the unitarity triangle fit. 

This conclusion receives corroboration by the observation that even without using $B\to\tau\nu$ at all, and using as input only $\epsilon_K$, $\Delta M_{B_s}/\Delta M_{B_d}$ and $|V_{cb}|$ (see Fig.~\ref{fig:tabsin2beta}), the predicted value of $\sin(2\beta)$ deviates by $1.8\sigma$ from its measurement (in this case we find $\sin(2\beta)^{\rm fit} = 0.814 \pm 0.081$). Thus,  possible new physics in $B\to\tau\nu$ can alleviate but not remove completely the tension in the fit.

We recall that the fit above is actually  the simple fit we had reported some time ago (now with updated lattice inputs) with its resulting  $\approx$ 2 $\sigma$ deviation~\cite{Lunghi:2008aa}. This fit is somewhat special as primarily one is only using $\Delta F = 2$ box graphs from $\epsilon_K$ and $\Delta M_{B_s}/\Delta M_{B_d}$ in conjunction with lattice inputs for $B_K$ and the SU(3) breaking ratio $\xi$. The experimental input from box graphs is clearly short-distance dominated and for the lattice these two inputs are particularly simple to calculate as the relevant 4-quark operators have no mixing with lower dimensional operators and also require no momentum injection. The prospects for further improvements in these calculations are high and the method should continue to provide an accurate and clean ``prediction" for $\sin (2 \beta)$ in the SM. So even if the current tensions get resolved, this type of fit should remain a viable way to test the SM as lattice calculations and experimental inputs continue to improve.
\begin{figure}[t]
\begin{center}
\includegraphics[width=0.6 \linewidth]{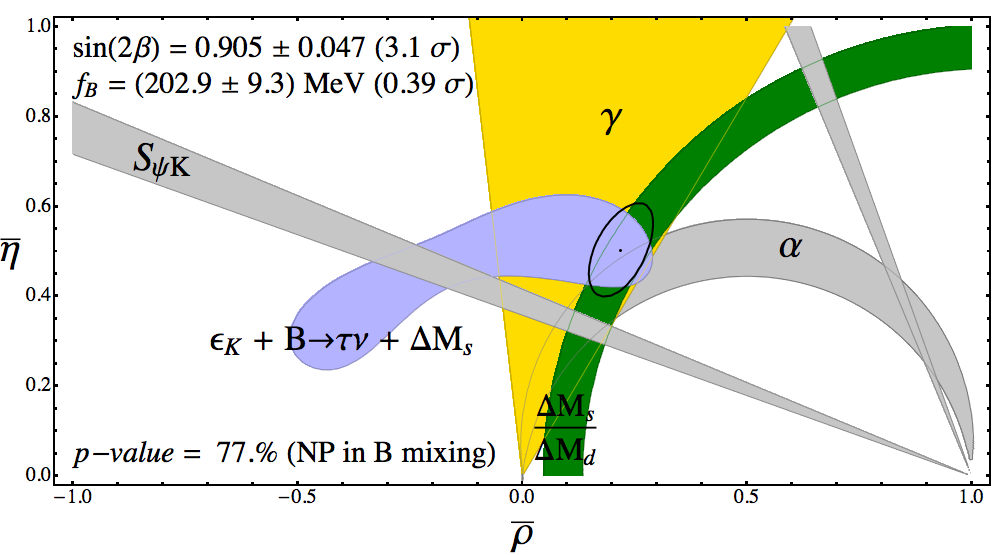}
\includegraphics[width=0.6 \linewidth]{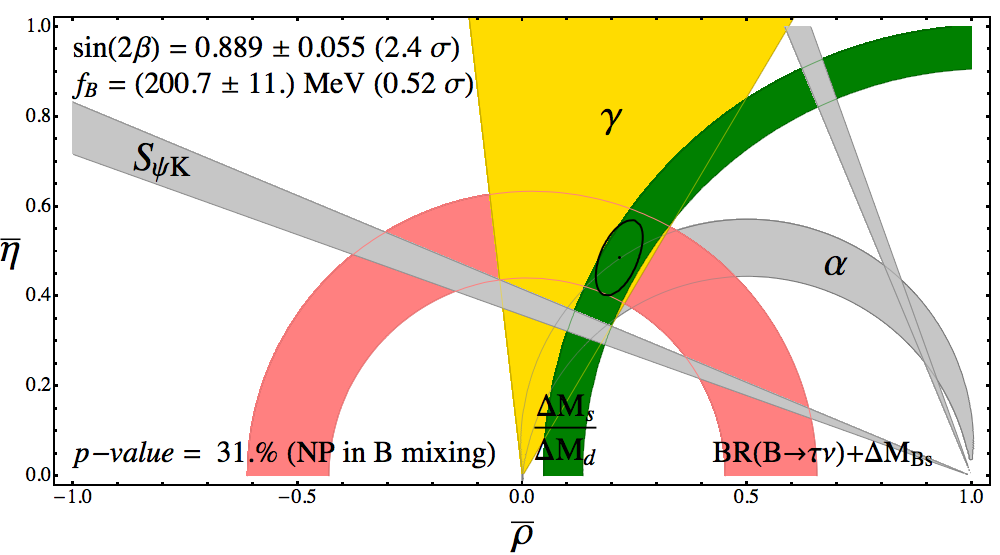}
\caption{Unitarity triangle fit without semileptonic decays (upper panel) and without use of $K$ mixing (lower panel). See the caption in Fig.~\ref{fig:utfit}. \hfill \label{extrafits}}
\end{center}
\end{figure}

\subsection{ Roles of $V_{cb}$, $\varepsilon_K$, $V_{ub}$ and of hadronic uncertainties.} 
The fit described above does use $V_{cb}$ where again the inclusive and exclusive methods differ mildly (about 1.7$\sigma$). Of greater concern here is that $\epsilon_K$ scales as $|V_{cb}|^4$ and therefore is very sensitive to the error on $V_{cb}$. We address this in two ways. First in the upper panel of Fig.~\ref{extrafits} we study a fit wherein no semi-leptonic input from $b \to c$ or $b \to u$ is being used. Instead, in this fit ${\rm BR} (B \to \tau \nu)$ and $\Delta M_{B_s}$ along with $\epsilon_K$, $\Delta M_{B_s}/\Delta M_{B_d}$ and $\gamma$ are used. Interestingly this fit gives
\begin{align}
\sin(2\beta)^{\rm fit} &= 0.905 \pm 0.047 \; , \label{no_vcb} \\
f_{B_d}^{\rm fit} &= (202.9 \pm 9.3)\; {\rm MeV} 
\end{align}
Thus, once again, $\sin (2 \beta)$ is off by $3.1\sigma$ whereas $f_{B_d}$ is in very good agreement with directly measured value which we again take to mean that the {\it bulk} of the discrepancy is in $\sin (2 \beta)$ rather than in $B \to \tau \nu$ or in $V_{cb}$.

Next we investigate the role of $\epsilon_K$. In the bottom panel of Fig.~\ref{extrafits}  we show a fit where only input from B-physics, namely $\Delta M_{B_s}/\Delta M_{B_d}$, $\Delta M_{B_s}$, $\gamma$, $|V_{cb}|$ and ${\rm BR} (B \to \tau \nu)$ are used. This fit yields,
\begin{align}
\sin(2\beta)^{\rm fit} &= 0.889 \pm 0.055 \; , \label{no_epsilonk} \\
f_{B_d}^{\rm fit} &= (200.7 \pm 11)\; {\rm MeV} 
\end{align}
Thus, $\sin(2\beta)^{\rm fit}$ is off by $\approx 2.4 \sigma$ and again $f_{B_d}^{\rm fit}$ is in good agreement with its direct determination. We are, therefore, led to conclude that the role of $\epsilon_K$ in the discrepancy is subdominant and that the bulk of the new physics contribution is likely to be in B--physics. As before, the fact that the fitted value of $f_{B_d}$ is in good agreement with its direct determination seems to suggest that the input ${\rm BR}(B\to \tau \nu)$ is most likely not in any large conflict with the SM, though, obviously we cannot rule out the possibility of it receiving a sub-dominant contribution from new physics.
\begin{figure}[t]
\begin{center}
\includegraphics[width=0.6 \linewidth]{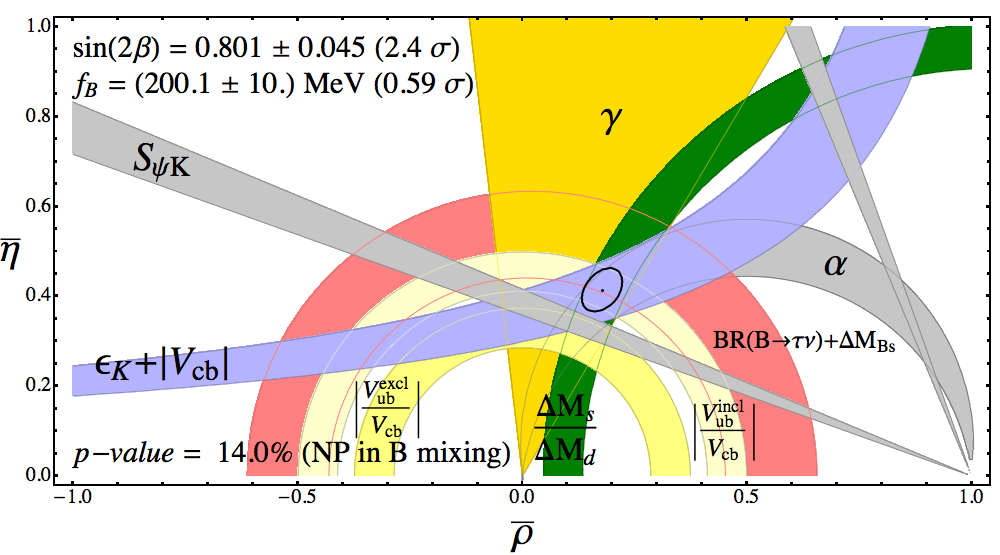}
\includegraphics[width=0.6 \linewidth]{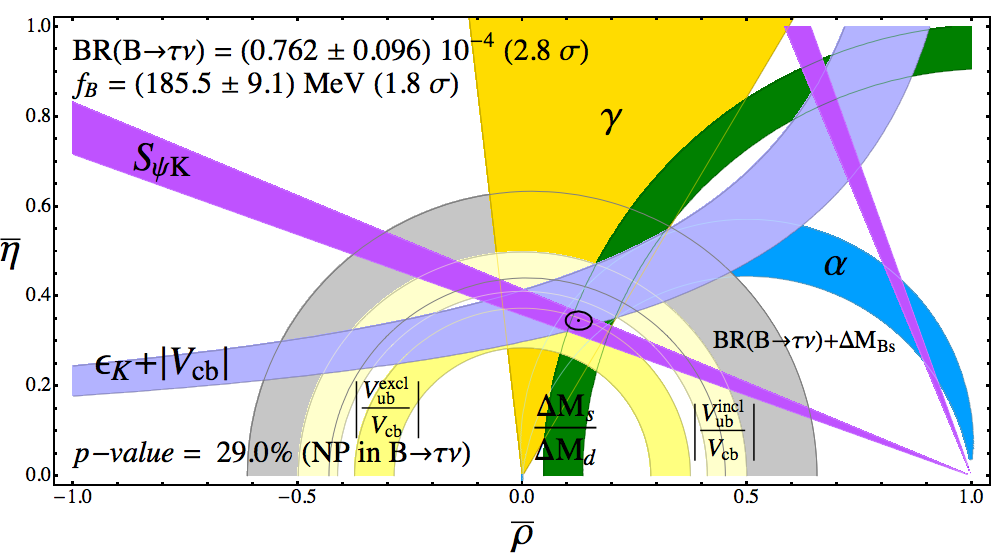}
\caption{Unitarity triangle fit with $V_{ub}$. We plot separately the constraints from inclusive and exclusive semileptonic $B$ decays. The contour, $p$-value and fit predictions are obtained using the $|V_{ub}|_{\rm tot}$. See the caption in Fig.~\ref{fig:utfit}. \hfill  \label{fig:utfit_vub}}
\end{center}
\end{figure}
\begin{figure}
\begin{center}
\includegraphics[width=0.6 \linewidth]{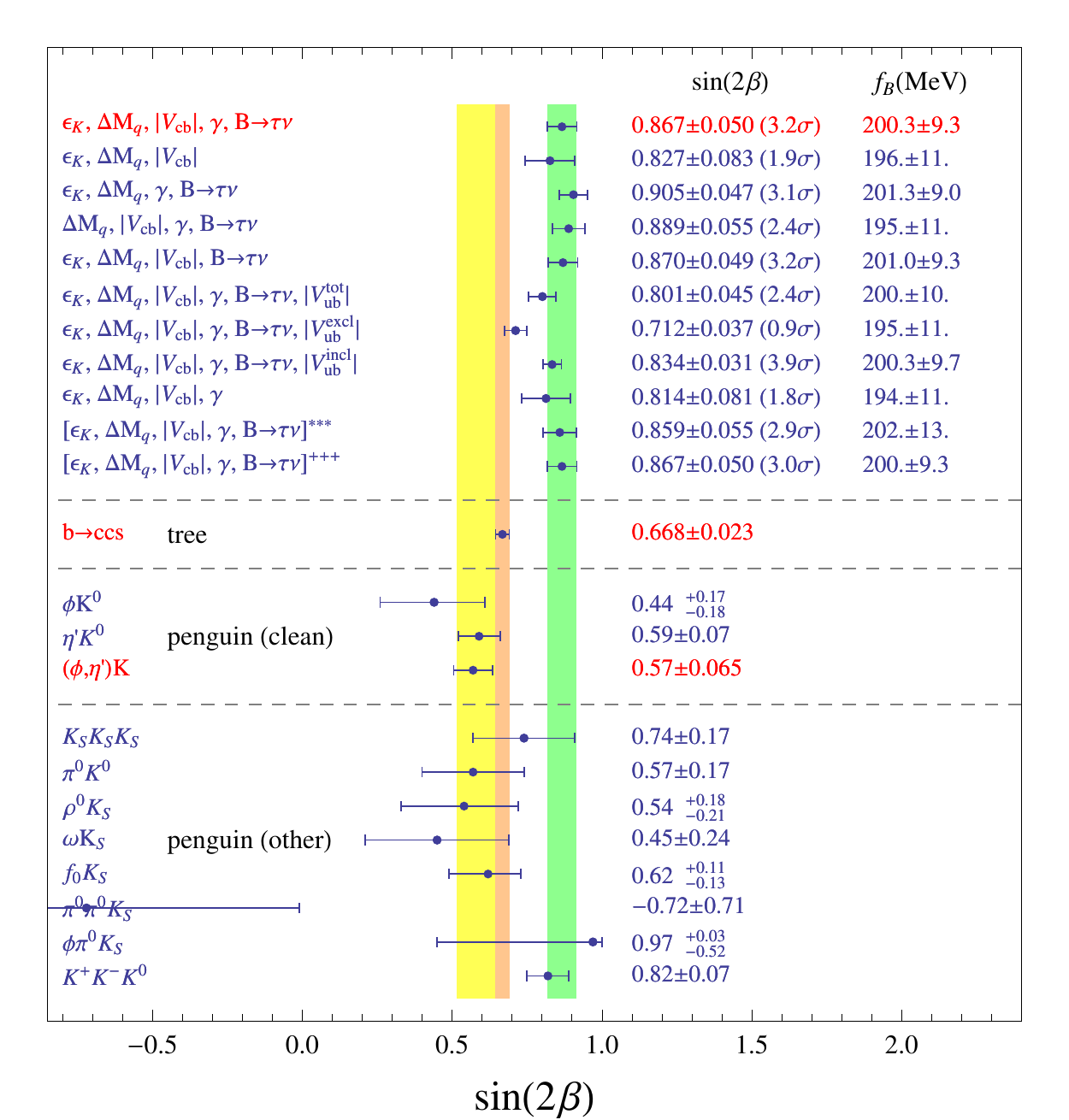}
\caption{Summary of $\sin(2\beta)$ determinations. The entry marked *** (tenth from the top) is obtained with lattice errors increased by 50\% over those given in Table~\ref{tab:utinputs} for each of the input quantities that we use and
the entry marked +++ (eleventh from the top) corresponds to adding an hadronic uncertainty $\delta \Delta S_{\psi K} = 0.021$ to the relation between $\sin (2\beta)$ and $S_{\psi K}$. See the text for further explanations. \hfill\label{fig:tabsin2beta}}
\end{center}
\end{figure}

For completeness, we present in Fig.~\ref{fig:utfit_vub} the results we obtain when including $V_{ub}$ in the fit. Note that inclusive and exclusive determinations of $|V_{ub}|$ differ at $3.3\sigma$ (see Table~\ref{tab:utinputs}) and, for this reason, are presented separately in the plot. Before taking the average, we add a 10\% model uncertainty to the inclusive determination. This reduces the discrepancy to $2.1\sigma$. We finally rescale the error on the average by the square root of the reduced chi-square (following the PDG recipe). In Table~\ref{tab:utinputs} we report the result we obtain and that we use in the fit.

A compilation of all the eleven fits that we studied for $\sin 2 \beta$ are shown in Fig.~\ref{fig:tabsin2beta}. Notice that there is only one case in here (8th from the top) where the discrepancy in $\sin 2 \beta$ is only $O(1\sigma)$. We believe this is primarily a reflection of the large ($\approx 14.4\%$) uncertainty with our combined $V_{ub}$ fit originating from the large disparity between inclusive and exclusive determinations. This is again a reminder of the fact that till this discrepancy gets removed, we cannot use $V_{ub}$ to draw any reliable  conclusion.
\begin{figure}
\begin{center}
\includegraphics[width=0.6 \linewidth]{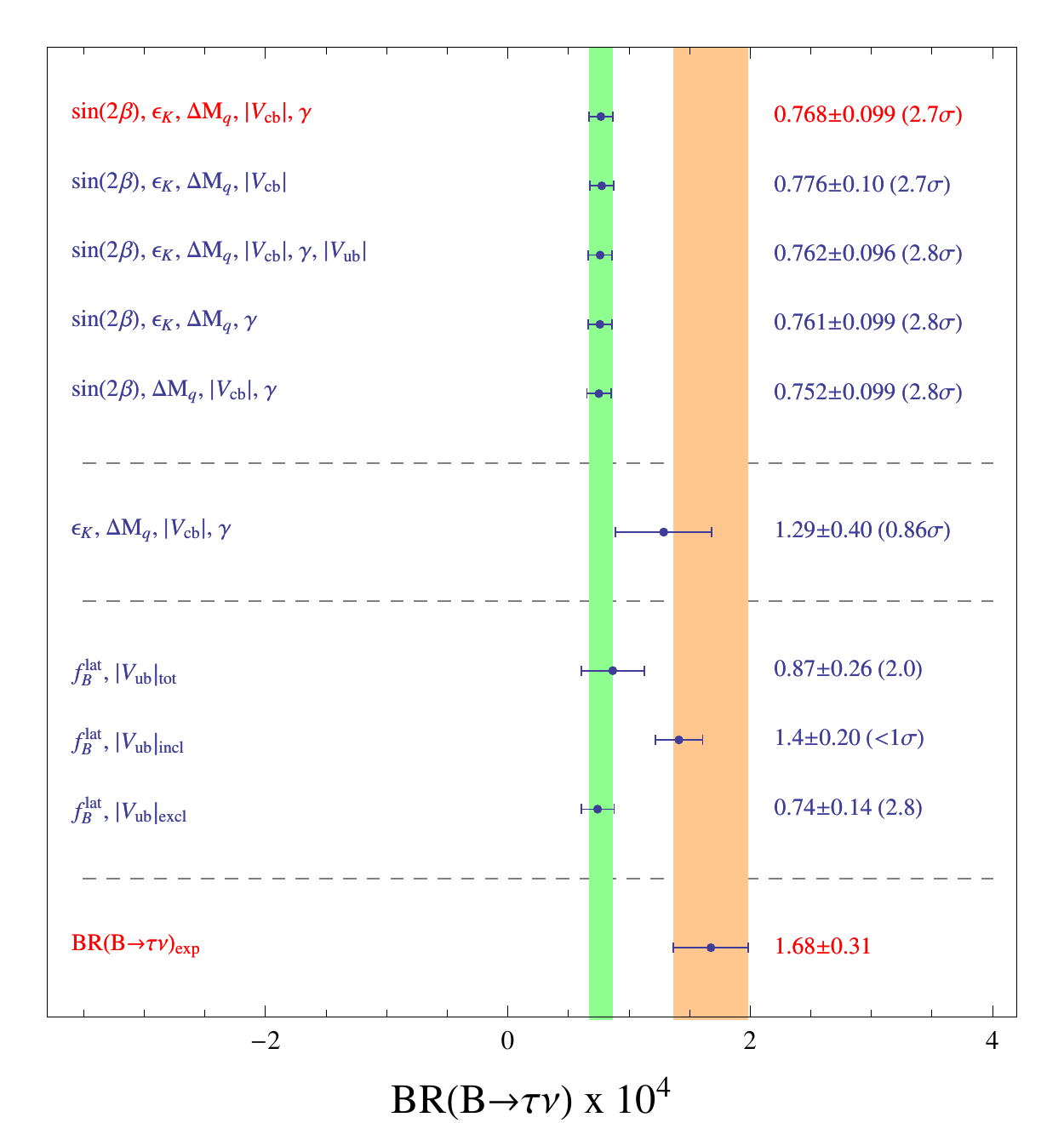}
\caption{Summary of  ${\rm BR} (B\to\tau\nu)$ determinations. \hfill \label{fig:tabbtn}}
\end{center}
\end{figure}

\section{\bf $B \to \tau  \nu$ and new physics}
Now with regard to $B \to \tau \nu$, Fig.~\ref{fig:tabbtn} shows a summary of predictions versus the measured ${\rm  BR}$. Notice that whenever  the measured value of $\sin ( 2 \beta)$ is used as an  input, the predicted BR is $\approx 2.7 \sigma$ from the measured one. In the preceding discussion we have emphasized that this seems to us to be a consequence of new physics largely in $B$ mixings and/or in $B_d \to \psi K_s$ decay. This conclusion receives further strong support when we try determine the $B\to\tau\nu$ branching ratio without using $\sin 2 \beta$. Indeed as shown in Fig.~\ref{fig:tabbtn} when we use $\epsilon_K$, $\Delta M_{B_q}$, $V_{cb}$ and $\gamma$ only, the fitted value of ${\rm BR}(B \to \tau \nu)$ is in very good agreement with the measured value. 

In principle, of course, the prediction for ${\rm  BR}(B \to \tau \nu)$ only needs the values of $f_{B_d}$ and of $V_{ub}$. Fixing now $f_{B_d} = 208 \pm 8~{\rm MeV}$ as  directly determined on the lattice (see Table~\ref{tab:utinputs}) we show the corresponding two predictions for the BR using separately the values of $V_{ub}$ determined in inclusive and in exclusive decays. It is clear that the inclusive determination yields results that are within one $\sigma$ of experiment (see also Fig.~\ref{fig:utfit}); however with $V_{ub}$ from exclusive modes (that makes use of the semileptonic form factor as determined on the lattice), the BR deviates by $\approx 2.8 \sigma$ from experiment. This may be a hint that lattice based exclusive methods have some intrinsic difficulty or that the exclusive modes are sensitive to some new physics that the inclusive modes are insensitive to, {\it e.g.} right-handed currents~\cite{Crivellin:2009sd,Buras:2010pz}. In either case, this reasoning suggests that we  try using  the value of  $V_{ub}$ given by inclusive methods only in our fit for determining $\sin 2 \beta$. This line of reasoning is also supported by the analysis presented in Ref.~\cite{Khodjamirian:2011ub} in which the discrepancy between the experimental determination and the SM prediction of ratio $R_{s/l}= {\rm BR} (B\to \pi\ell\nu)/{\rm BR} (B\to \tau\nu)$ is considered. Note that the authors of Ref.~\cite{Khodjamirian:2011ub} find that the experimental value of this ratio is about a factor of 2 smaller than the SM prediction and that this discrepancy is independent of whether lattice QCD or Light-Cone QCD Sum Rules are used to determine the $B\to \pi$ form factor and the $B$ decay constant. This result can be seen as a solid consistency check of the lattice QCD calculation of the $B\to \pi$ form factor. Within the SM this ratio is independent of short distance physics (the $|V_{ub}|^2$ factors cancel out) and measures the ratio of the $B\to\pi$ form factor to the $B$ decay constant. New physics in right--handed currents affects differently the $B\to\pi\ell\nu$ and $B\to\tau\nu$ transitions and might be responsible for the observed discrepancy. 

\section{\bf Summary of fits, perspective \& outlook.}
The result of our analysis strongly suggests that the SM predicted value of $\sin ( 2 \beta)$ is  around 0.85 whereas the value measured experimentally via the gold plated $\psi K_s$ mode is around 0.66 constituting a deviation of about 3$\sigma$ from the SM (see Fig.~\ref{fig:tabsin2beta}). To put this result in a broader perspective let us now recall that in fact in the SM $\sin (2 \beta)$ can also be measured via the penguin dominated modes (see Fig.~\ref{fig:tabsin2beta})~\cite{Grossman:1996ke,Fleischer:1996bv,Grossman:1997gr,London:1997zk}. Unfortunately several of these modes suffer from a potentially large tree pollution, though there are good reasons to believe that the $\eta^{\prime} K_s$, $\phi K_s$ and 3 $K_s$ modes are rather clean~\cite{Cheng:2005bg,Cheng:2005ug,Beneke:2005pu} wherein the deviations from $\sin 2 \beta$ are expected to be only O(few \%). The striking aspect of these three clean modes as well as many others penguin dominated modes (see Fig.~\ref{fig:tabsin2beta}) is that the central values of almost all of them tend to be even smaller than the value (0.66), measured in $\psi K_s$, and consequently tend to exhibit even a larger deviation from the SM prediction of around 0.85. Thus, seen  in the light of our analysis, the deviation in these penguin modes suggests the presence of new CP-violating physics not just in $B$-mixing but also in $b \to s$ penguin transitions.

 Moreover, the large difference ($\approx (14.4 \pm 2.9) \%$)~\cite{PDG_10} in the direct CP asymmetry  measured in $B^0  \to K^+ \pi^-$  versus that in $B^+ \to K^+ \pi^0$ provides another hint that $b \to s$ penguin transitions may be receiving the contribution from a beyond the SM source of CP-violation (for alternate explanation see Refs.~\cite{Mishima:2011qy,Gronau:2006ha,Cheng:2009cn}). To briefly recapitulate, in the SM one naively expects this difference to be vanishingly small and careful estimates based on QCD factorization ideas suggest that it is very difficult to get a difference much larger than $(2.2 \pm 2.4) \%$~\cite{Lunghi:2009sm}.

Of course, if $b \to s$ penguin transitions ($\Delta Flavor =1$)  are receiving contributions from new physics, then it is quite unnatural  for $B_s$ mixing amplitudes ($\Delta Flavor =2$) to remain unaffected. Therefore, this reasoning suggests that we should expect non-vanishing CP asymmetries in $B_s \to \psi \phi$ as well as a non-vanishing  di-lepton asymmetry in $B_s \to X_s l \nu$.  As is well known,  at Fermilab, in the past couple of years CDF and D0 experiments have been studying CP asymmetry in $B_s \to \psi \phi$. The latest result with about 6 ${\rm fb}^{-1}$ from each experiment seems to reveal a reduction from $\sim \!\! 1.8 \sigma$ tension to $\sim \!\! 1  \sigma$ from the SM~\cite{RK_ICHEP10,DT_BF10,HFAG10}. Thus, findings in $B_s \to \psi \phi$ from Fermilab and from LHCb are eagerly awaited.

Another interesting and potentially very important development with regard to non-standard CP in $B_s$ is that last year D0  announced the observation of a large dimuon asymmetry in B-decays amounting to a deviation of ($\approx 3.2 \sigma$) from the minuscule asymmetry predicted in the SM~\cite{Abazov:2010hv,Abazov:2010hj}. They attribute this largely to originate from $B_s$ mixing. While this is a very exciting development, their experimental analysis is extremely challenging and a confirmation is highly desirable before their findings can be safely assumed. Note, though, HFAG~\cite{HFAG10} has combined CDF and D0 results on $B_s \to \psi \phi$
and on the  dimuon asymmetry, $A_{sl}^s$ and finds the deviation 
from the SM to be around 2.7$\sigma$.

Be that as it may, we reiterate that our analysis  suggests that the deviation from the SM in $\sin (2 \beta)$ is difficult to reconcile with errors in the inputs from the  lattice  that we use,  and strongly suggests the presence of a non-standard source of CP violation largely in $B$/$B_s$ mixings, thereby predicting that non-standard signals of CP violation in $S (B_d \to \eta^{\prime} K_s, \phi K_s, 3 K_s etc. )$ as well as in $S(B_s \to \psi \phi)$, and  the semileptonic  and di-lepton asymmetries  in $B_s$, and possibly also in $B_d$, decays will persist and survive further scrutiny in experiments at the intensity frontiers such as Fermilab (CDF, D0), LHCb and the Super-B factories. Lastly, the fact that our analysis rules out the possibility that new physics exclusively  in kaon mixing is responsible for the deviations in $\sin (2 \beta)$,  has the very important  repercussions for the  mass scale of the underlying new physics contributing to  these deviations: model independent analysis then  imply that the relevant mass scale of the  new physics is necessarily relatively low, {\it i.e.} below O(2 TeV)~\cite{Lunghi:2009sm,no_LR}. Thus, collider experiments at the high energy frontier at LHC and possibly even at Fermilab should see direct signals of the  underlying degrees of freedom appearing in any relevant beyond the Standard Model scenario. 

\section{Aftermath: BSM possibilities} 
Let us next discuss  a model independent point of view as to the possible underlying cause for these anomalies and then two specific models that may  be relevant.

\subsection{Brief Summary of the model independent analysis}
One of the important issue is how these B-CP anomalies will impact search for New Physics at the LHC wherein a knowledge of  underlying scale of NP would be very useful. With this in mind we~\cite{Lunghi:2009sm} write down dimension-6 operators under the general assumptions of NP in $\Delta Flavor=2$ effective Hamiltonian for K, $B_d$ or $B_s$ mixing or for the case of $\Delta Flavor=1$ Hamiltonian that may be relevant for $b \to s$ penguin transitions~\cite{Lunghi:2009sm}. Our model independent analysis shows that the scale of CP violating NP is only a few hundred GeV if it originates from $b \to s$, $\Delta Flavor=1$ penguin Hamiltonian. It rises to about a few TeV if it originates from $B_d$ and/or $B_s$ mixing. From the perspective of LHC the scenario that is the most pessimistic, with NP scale  in the range of a few tens of TeVs, is when all of the NP resides only in the dimension-6 LR-operator relevant for the $K -\bar K$ mixing~\cite{Beall:1981ze}. However, in the preceding sections we have shown that bulk of the deviation from the SM does not originate in $\epsilon_K$ or $K-\bar K$ mixing. This, therefore, has the important consequence that taken seriously these discrepancies with the CKM hint at scale of new physics that is quite likely to be less  than a few TeV, possibly even a few hundred GeV. Note also that in our 08-09 work~\cite{Lunghi:2009sm} we were unable to rule out the possibility that dominantly NP resides in $K$--mixing; this only became possible in our 2010 analysis~\cite{Lunghi:2010gv}.

\subsection{Warped Flavordynamics \& duality}
Perhaps the most interesting and even compelling BSM scenario is that of warped extra dimensional models~\cite{RS99} as it offers a simultaneous resolution to EW-Planck hierarchy as well as flavor puzzle~\cite{GN99,GP00}. While explicit flavor models are still evolving, potentially this class of models has many interesting features; for example, in general one expects several new BSM CP-odd phases (presumably O(1))~\cite{APS1} that can have important repercussions for flavor physics. Indeed in the simplest scenario it was {\it predicted}~\cite{APS2} that there should be smallish ({\it i.e.} O(20\%)) deviations from the SM  in $B_d$ decays to penguin-dominated final states such as $\phi K_s$,  $\eta' K_s$ etc as well as the possibility of a largish CP-odd phase in $B_s$ mixing which then of course has manifestations in {\it e.g.} $B_s \to  \psi \phi$, time-dependent asymmetries in $B_d \to (\rho ,K^*) \gamma$, etc~\cite{APS1,APS2}. In fact there was also a mild CP problem in that very simple rough estimates suggest neutron-EDM should  be bigger than current bounds~\cite{Baker:2006ts}  by about an order of magnitude.
 
Note though that in this original study, for simplicity, it was assumed that $B_d$ mixing was essentially described by the SM. More recently there have been two extensive studies of the possibility of warped models being the origin of the several hints in B, $B_s$ decays mentioned above~\cite{AJB081,AJB082,MN08}.

A common feature of  these warped models is that they also imply the existence of various Kaluza-Klein states, excited counterparts of the gluon, weak gauge bosons and of  the graviton with masses heavier than about 3 TeV~\cite{adms03}. Note also that unless  the masses of these particles are less than about 3 TeV their direct detection at LHC will be very difficult~\cite{ABKP06,LR_KKG07,shri1,shri2,LR07,ADPS07,AAS08,AS08bis,APS06}.

Since viable explicit models are still being developed, it is useful
to emphasize generic predictions of warped flavor models.
Top quark should exhibit large flavor violations via {\it e.g.}
$t \to c(u) Z$~\cite{APS06}, largish $D^0$ mixing with possibly enhanced CP violations, beyond the SM CP asymmetries are also possible in
$B_d$, $B_s$ physics, also polarized top quarks endowed with some
forward-backward asymmetry~\cite{mainz} should be expected. Furthermore, KK particles such as the KK-glue, graviton, Z' should have large
BR to top quark pairs and in fact the tops should be boosted
since the decaying resonances are expected to have TeV-scale masses~\cite{ABKP06,LR_KKG07,ADPS07,shri1,shri2}.
  
 An extremely interesting subtlety about these 5-dimensional warped models is that they are supposed to be dual to some 4-dimensional models with strong dynamics~\cite{JMM97,Gubser:1998bc,Witten:1998qj,NIMA00,Rattazzi:2000hs,Contino03,Agashe04}. This serves as motivation to search for effective 4-dimensional models that provide a good description of the data.

\subsection{Extension of SM to four generations: SM4 }
SM with four generations provides a rather simple and an interesting extension to address the B-CP anomalies~\cite{AS08,AS10,Nandi:2010zx,AJB10,AJB10_D0,GH10,AL091,AL092}. Actually,for several reasons, SM4 is of considerable interest irrespective of these anomalies: \\

\noindent $\bullet$ The heavier quarks could form condensates and thereby play an important role in dynamical electroweak symmetry breaking~\cite{PQ,MH09,BH06,BURD07}. \\

\noindent $\bullet$ Two new CP -odd phases  and the heavier quark masses also significantly alleviate one of the key difficulty that SM3 has for baryogenesis~\cite{GH08}. \\

\noindent $\bullet$ SM4 can open new avenues for addressing the dark matter 
issue~\cite{GEV,LLS}. \\

\noindent $\bullet$ Besides, since we have already seen three families, it is natural to ask
why not the fourth?\\

Note also that while practically all BSM scenarios have difficult time
accounting for the (almost) absence of FCNC (in processes such as
${\rm BR} ( B \to X_s \gamma)$), SM4 explains this readily. First of all in SM4 (as in SM) FCNC are loop suppressed via the GIM~\cite{Glashow:1970gm} mechanism. Furthermore, as you
extend the $3\times 3$  matrix to $4\times 4$ and impose unitarity, the hierarchical structure of CKM matrix extended to $4\times 4$ automatically allows 
only small residual CP-conserving effects in quantities such as
${\rm BR}(B \to X_s \gamma)$; on the other hand, there can be dramatic difference in CP violating observables where in the SM null results are predicted~\cite{GS06}.

In contrast to CP-conserving FCNC, since there are now two new
CP-odd phases, they can cause large (O(1)) CP -asymmetries in
channels that the CKM phase has negligible effect in the SM~\cite{GS06}.
This is expected to be the case {\it e.g.} in Bs mixing,
causing S($B_s \to \psi \phi$) and the semileptonic asymmetry,
$a_{sl}^s$, in $B_s \to X_s \ell\nu$
to be non-vanishing. Similarly there may be non-vanishing 
CP-asymmetries in
 $b \to s \gamma$, $b \to s l l$, $B_d \to \eta' K_s$,
 $B_d \to \phi K_s$ wherein the SM one expects negligible effects. 
 Moreover, we should also expect the effects due to  
 an additional phase in $B_d$ mixing beyond what's there in the SM.
 This manifests itself say as a deviation in {\it e.g}
 $\sin 2\beta$
 from the SM predicted value and also can cause the
 semi-leptonic asymmetry, $a_{sl}^d$ to differ from the SM predicted
 value (that is negligibly small).
 
It may be useful also to note that seen from the perspectives of SM4
the hierarchy puzzle may just be a historical accident.

Natural mass scale for a Higgs particle in SM4, where the heavy quarks
are geared towards EW symmetry breaking, is 
around $2 m_{b'}$. Such a heavy Higgs would of course have
very clean decays to  $H \to ZZ$.
  
Electroweak precision tests do not rule out the existence of a 4th generation though they restrict the mass splitting between the 4th generation doublet $(t^\prime, b^\prime)$ of quarks to be less than around 75 GeV~\cite{Tait,Le}. This requires some 10\% degeneracy in their masses.
One cautionary remark is that such heavy quark masses means
large Yukawa couplings, therefore many perturbative calculations,
including those relevant for EW
precision tests may receive large corrections.

Furthermore LEP experiments require that the 4th generation neutral lepton has to be rather heavy $\gtrsim m_Z/2$; this begs the question as to why there should be such a huge disparity with the three almost massless neutrinos of the conventional SM3~\cite{GEV,LLS}. Issues such as these are interesting and require further investigations.

\begin{figure}
\begin{center}
\includegraphics[width=0.6 \linewidth]{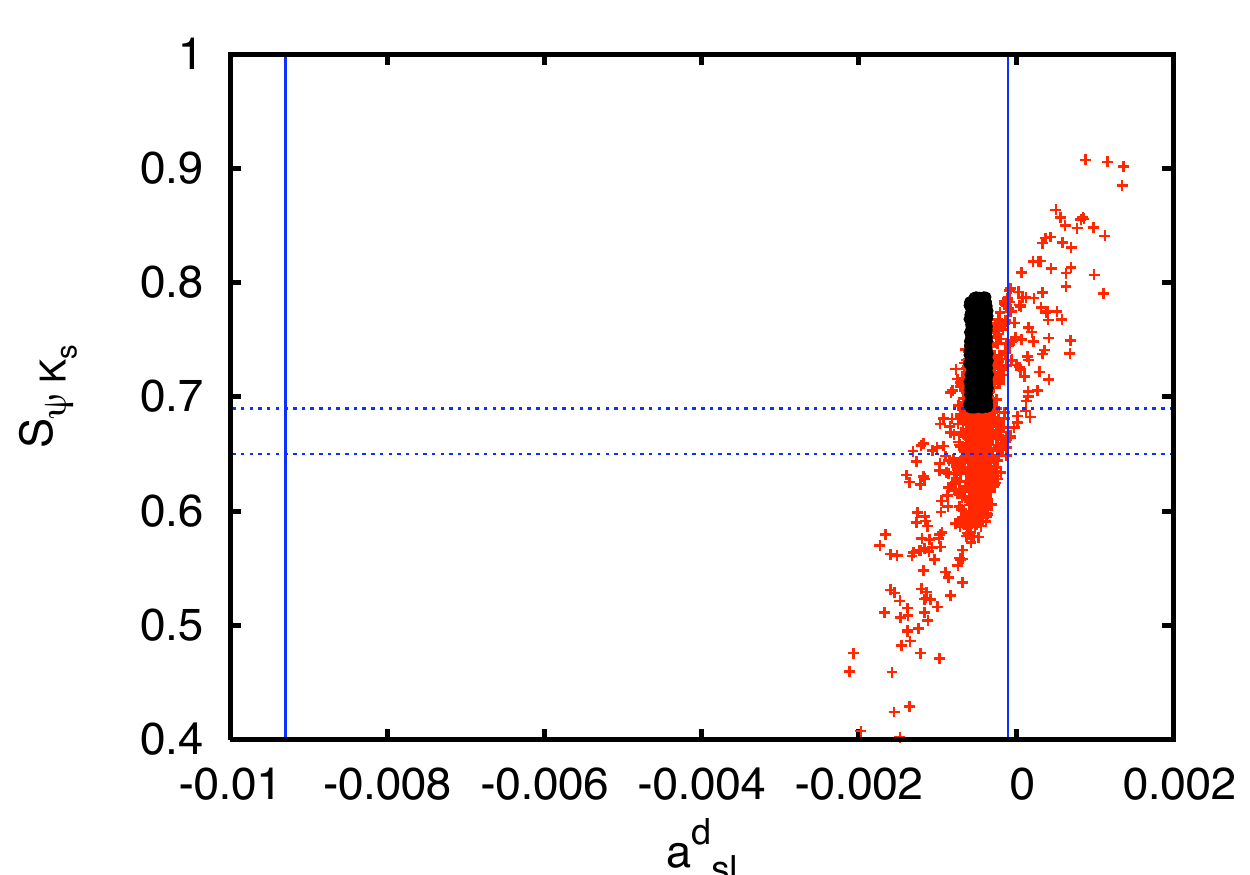}
\caption{Correlation between $S{\psi K_S}$ and the semi-leptonic asymmetry, $a_{sl}^d$ in SM4 is shown for $m_{t'}$ varying between 375 and 575 GeV. The experimentally allowed region as well as the SM3 
bounds are also shown. See Ref.~\cite{Nandi:2010zx} for details. \hfill  \label{fig:asld_psiks}}
\end{center}
\end{figure}

We will briefly now summarize the possible relevance of SM4 to alleviating the discrepancies in the CKM picture that has been our primary focus here. In particular, the tension in S($\psi K_s$) that we emphasized has been examined in~\cite{Nandi:2010zx}. Fig~\ref{fig:asld_psiks} from Ref.~\cite{Nandi:2010zx} shows a study of the correlation between the predicted value of $\sin 2 \beta$ and $a_{sl}^d$, which is the semi-leptonic asymmetry in $B_d \to X_d l \nu$. In this study all known experimental constraints such as $\epsilon_K$, $Br(K^+ \to \pi^+ \nu \bar \nu)$, $\Delta M_d$, $\Delta M_s$, constraints on the unitarity angle $\gamma$ etc have all been incorporated. As is evident from the figure, while the SM4 can accommodate the measured value of S($\psi K_s$) it also requires simultaneously that $a_{sl}^d \gtrsim -0.001$, which is only about a factor of a few different from the SM3. This underscores  another attractive aspect of SM4: it is rather predictive, highly constrained extension of the SM and can be ruled out with relative ease. In this specific illustration it requires improved determination of $a_{sl}^d$ as well as $S_{\psi K_s}$ both theoretically and experimentally. Improved bound from the B factories, who need to update their several years old result~\cite{HFAG10} on $a_{sl}^d$ would be very useful.

\begin{figure}
\begin{center}
\includegraphics[width=0.6 \linewidth]{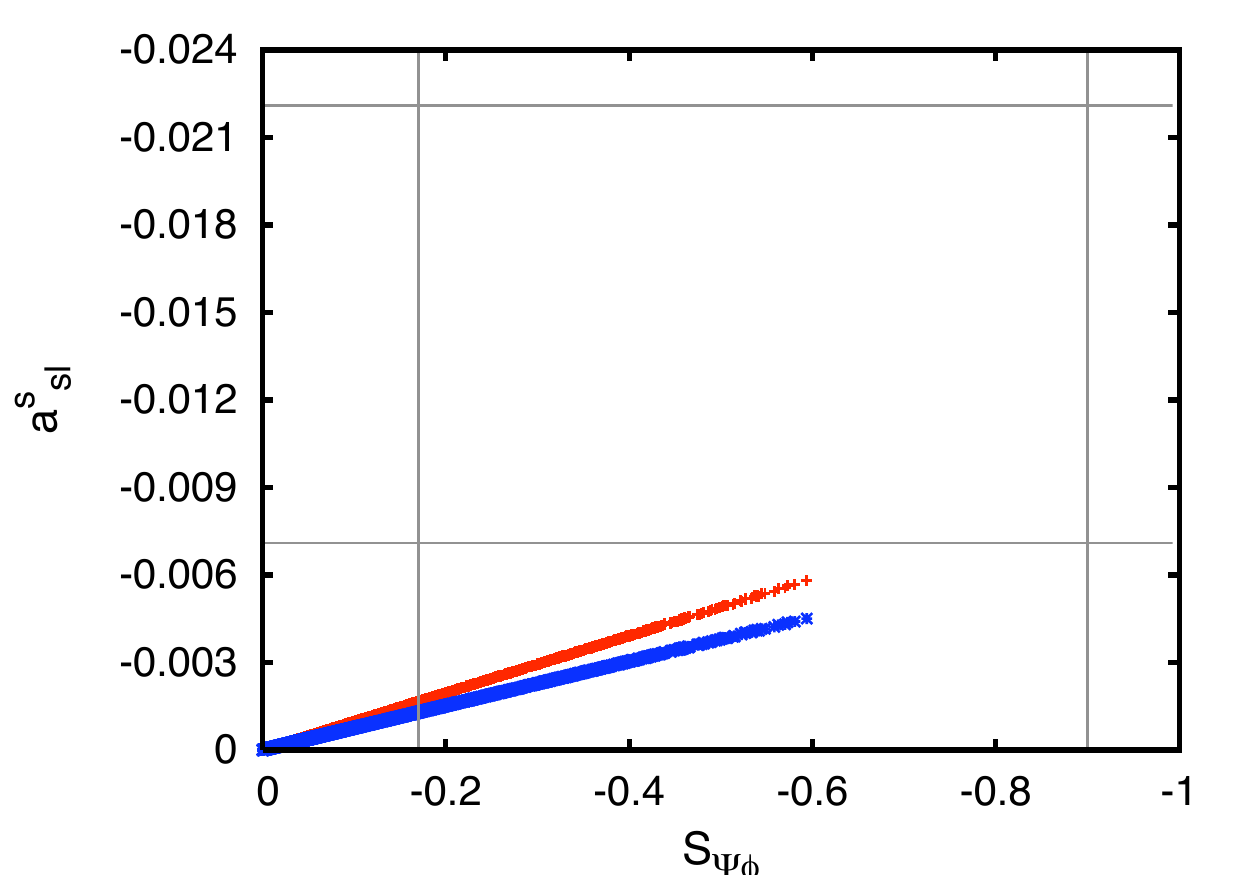}
\caption{Correlation between $S{\psi K_S}$ and the semi-leptonic asymmetry, $a_{sl}^d$ in SM4 is shown for $m_{t'}$ varying between 375 and 575 GeV. The experimentally allowed region ($1\sigma$) is shown. See Ref.~\cite{Nandi:2010zx} for details.   \label{fig:asls_psiphi}}
\end{center}
\end{figure}

Another interesting example is  the semi-leptonic asymmetry in $B_s$, $a_{sl}^s$ and its correlation with S($B_s \to \psi \phi$), see Fig.~\ref{fig:asls_psiphi} from ~\cite{Nandi:2010zx}. It is interesting to note here that while SM4 allows a much larger semi-leptonic asymmetry as shown, it is still not large enough to explain the central value of the asymmetry reported by the recent D0 result~\cite{Abazov:2010hj}. Thus, if improved experimental results uphold near the central value of the D0 experiment, then SM4 may also be ruled out.

Finally, let us also briefly mention that experimental searches for quarks ($t'$, $b'$) of the 4th generation have already been underway at Fermilab leading to a lower bound of around 350 GeV~\cite{CDF_bounds}. We should expect intensified searches at LHC especially since this is something that can be achieved even in the early 7 GeV run~\cite{AGS11}. It is expected that after several years of efforts, LHC should be able to find these quarks or put a bound close to a TeV~\cite{DW_NTU10}.

\section*{Acknowledgments}
We want to thank Jean-Marie Frere, Maurizio Pierini, Yuval Grossman, Uli Haisch and Alexander Khodjamirian for discussions and suggestions. This research was supported in part by the U.S. DOE contract No.DE-AC02-98CH10886(BNL).


\begin{thebibliography}{99}
\bibitem{BK_define}
 Recall $B_K \equiv < K^0 | [\bar s {\gamma_\mu}L d]^2|\bar K^0> /[(8/3) f_K^2 m_K^2]$ where $m_K$ and $f_K$ are the mass and the pseudoscalar decay constant of the kaon, in the convention that
experimentally $f_K \approx 160 MeV$.
\bibitem{thanks_Taku}
    We want to thank Taku Izubuchi for help with this figure.
\bibitem{footnote_Vub}
The inclusive and exclusive determinations of $|V_{ub}|$ differ at the $1.8\sigma$ level even after including an additional 10\% model uncertainty to the former. We first calculate the standard weighted average of these two determinations and then, following the PDG prescription, we rescale the resulting uncertainty by the square root of the reduced chi-square.
\bibitem{no_alpha}
In contrast to $\gamma$, $\alpha$ is not used as an input since it receives appreciable contribution from  penguin amplitudes which are sensitive to new physics.
\bibitem{no_gamma}
Note that when $\gamma$ is not used as an input and only $\epsilon_K$, $\Delta M_s/\Delta M_d$ and ${\rm BR} (B \to \tau \nu)$ are used, the deviation of the fitted $\sin (2 \beta)$ from the measured one stays unchanged at 3.2$\sigma$, see Fig.~\ref{fig:tabsin2beta}.
\end{thebibliography}
\end{document}